\documentclass[aps,prl,reprint,superscriptaddress]{revtex4-1}

\usepackage{graphicx}
\usepackage{color,amsmath}
\usepackage{amsfonts}
\usepackage[normalem]{ulem}
\usepackage{hyperref}
\usepackage{lipsum}

\begin{document}
\title{Relating Andreev Bound States and Supercurrents in Hybrid Josephson Junctions}

\author{F.~Nichele}
\email[email: ]{fni@ibm.zurich.com}
\affiliation{Center for Quantum Devices and Microsoft Quantum Lab - Copenhagen, Niels Bohr Institute, University of Copenhagen, Universitetsparken 5, 2100 Copenhagen, Denmark}
\affiliation{IBM Research - Zurich, S\"aumerstrasse 4, 8803 R\"uschlikon, Switzerland.}

\author{E.~Portol\'es}
\affiliation{Center for Quantum Devices and Microsoft Quantum Lab - Copenhagen, Niels Bohr Institute, University of Copenhagen, Universitetsparken 5, 2100 Copenhagen, Denmark}

\author{A.~Fornieri}
\affiliation{Center for Quantum Devices and Microsoft Quantum Lab - Copenhagen, Niels Bohr Institute, University of Copenhagen, Universitetsparken 5, 2100 Copenhagen, Denmark}

\author{A.~M.~Whiticar}
\affiliation{Center for Quantum Devices and Microsoft Quantum Lab - Copenhagen, Niels Bohr Institute, University of Copenhagen, Universitetsparken 5, 2100 Copenhagen, Denmark}

\author{A.~C.~C.~Drachmann}
\affiliation{Center for Quantum Devices and Microsoft Quantum Lab - Copenhagen, Niels Bohr Institute, University of Copenhagen, Universitetsparken 5, 2100 Copenhagen, Denmark}

\author{T.~Wang}
\affiliation{Department of Physics and Astronomy and Microsoft Quantum Materials Lab - Purdue, Purdue University, West Lafayette, Indiana 47907 USA}
\affiliation{Birck Nanotechnology Center, Purdue University, West Lafayette, Indiana 47907 USA}

\author{G.~C.~Gardner}
\affiliation{School of Materials Engineering, Purdue University, West Lafayette, Indiana 47907 USA}
\affiliation{Birck Nanotechnology Center, Purdue University, West Lafayette, Indiana 47907 USA}

\author{C. Thomas}
\affiliation{Department of Physics and Astronomy and Microsoft Quantum Materials Lab - Purdue, Purdue University, West Lafayette, Indiana 47907 USA}
\affiliation{Birck Nanotechnology Center, Purdue University, West Lafayette, Indiana 47907 USA}

\author{A.~T.~Hatke}
\affiliation{Department of Physics and Astronomy and Microsoft Quantum Materials Lab - Purdue, Purdue University, West Lafayette, Indiana 47907 USA}
\affiliation{Birck Nanotechnology Center, Purdue University, West Lafayette, Indiana 47907 USA}

\author{M.~J.~Manfra}
\affiliation{Department of Physics and Astronomy and Microsoft Quantum Materials Lab - Purdue, Purdue University, West Lafayette, Indiana 47907 USA}
\affiliation{Birck Nanotechnology Center, Purdue University, West Lafayette, Indiana 47907 USA}
\affiliation{School of Materials Engineering, Purdue University, West Lafayette, Indiana 47907 USA}
\affiliation{School of Electrical and Computer Engineering, Purdue University, West Lafayette, Indiana 47907 USA}

\author{C.~M.~Marcus}
\email[email: ]{marcus@nbi.ku.dk}
\affiliation{Center for Quantum Devices and Microsoft Quantum Lab - Copenhagen, Niels Bohr Institute, University of Copenhagen, Universitetsparken 5, 2100 Copenhagen, Denmark}

\date{\today}

\begin{abstract}
We investigate superconducting quantum interference devices consisting of two highly transmissive Josephson junctions coupled by a superconducting loop, all defined in an epitaxial InAs/Al heterostructure. A novel device design allows for independent measurements of the Andreev bound state spectrum within the normal region of a junction and the resulting current-phase relation. We show that knowledge of the Andreev bound state spectrum alone is enough to derive the independently measured phase dependent supercurrent. On the other hand, the opposite relation does not generally hold true as details of the energy spectrum are averaged out in a critical current measurement. Finally, quantitative understanding of field dependent spectrum and supercurrent require taking into account the second junction in the loop and the kinetic inductance of the epitaxial Al film.
\end{abstract}

\maketitle

\begin{figure*}
\includegraphics[width=2\columnwidth]{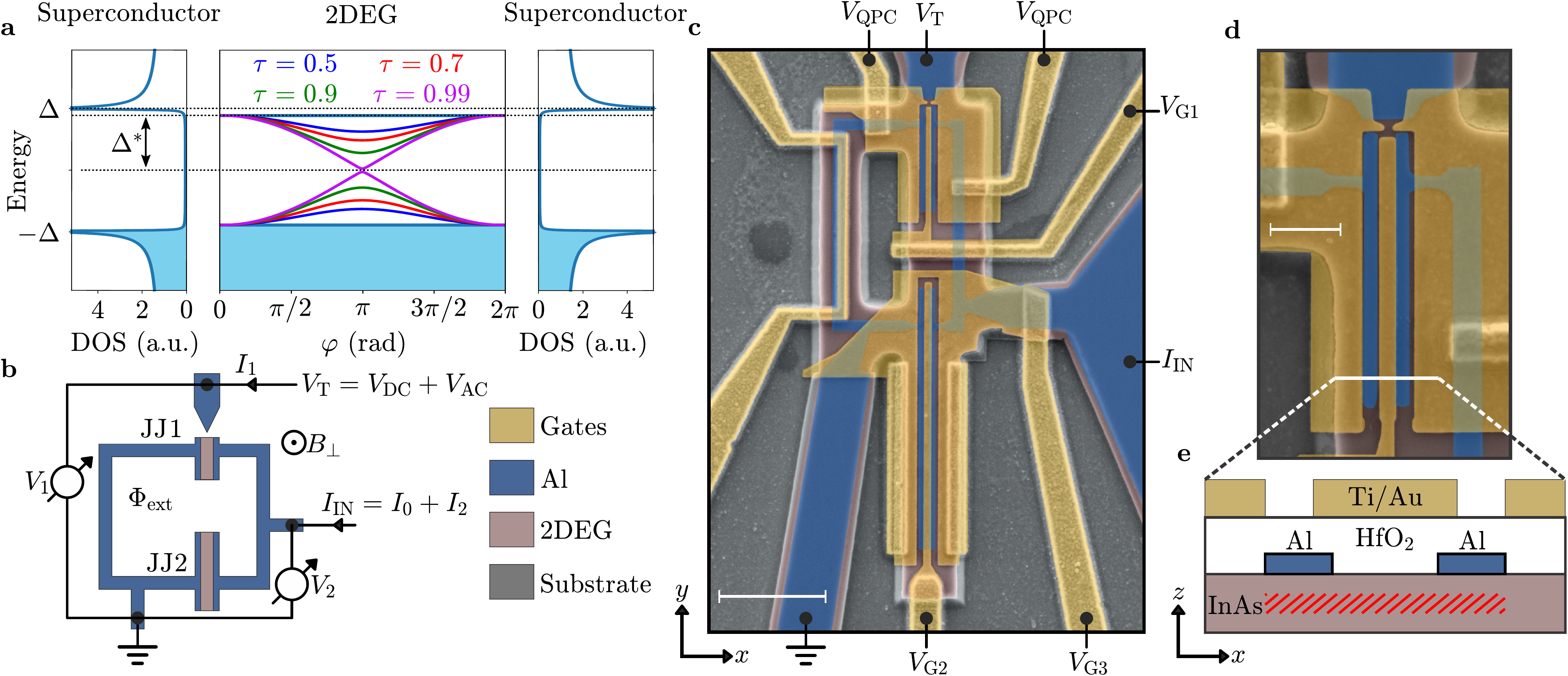}
\caption{(a) Schematic representation of the energy spectrum of a planar JJ. Two superconducting leads with a BCS-like DOS and induced gap $\Delta^*$ laterally confine a normal region in the 2DEG, in which discrete ABSs form within the induced superconducting gap. (b) Schematic representation of the device under study and its measurement setup. (c) False-color scanning electron micrograph of the device under study The insulating substrate is gray, the semiconductor red and the thin Al film blue. Top gates, colored yellow, were deposited over an insulating $\mathrm{HfO_2}$ layer (not visible). A metallic strip-line, not used in this work, was deposited on the left-hand side of the loop. Scale bar is $2~\mathrm{\mu m}$. (d) Zoom-in close to JJ1. A central top gate tuned the density in the normal region of JJ1, two additional top gates confined electrons below the superconducting leads and electrostatically defined a tunneling constriction between the normal region of JJ1 and a superconducting plane. Scale bar is $500~\mathrm{nm}$. (e) Schematic cross section of JJ1. A negative gate voltage applied to the lateral gates depleted lateral InAs regions, leaving a conducting electron layer solely below the Al contacts and within the normal region of JJ1.}
\label{fig:1}
\end{figure*}

Josephson junctions (JJs) in superconductor/semiconductor hybrids are objects of intense study. Electrostatic tuning of the critical current enables voltage controlled superconducting qubits~\cite{Larsen2015,Casparis2018,Wang2019}. Ballistic electron motion and spin-orbit interaction allow unique functionalities, such as spin-dependent supercurrents~\cite{Hart2016} and anomalous current-phase relation (CPR)~\cite{Szombati2016,Spanton2017,Assouline2019}. Most remarkably, Andreev bound state (ABS) manipulation via phase biasing and Zeeman fields might stabilize electronic phases with topological protection~\cite{Hell2017,Pientka2017,Fornieri2019,Ren2019}.
Several studies characterized hybrid JJs in terms of CPR~\cite{Golubov2004,DellaRocca2007,Bretheau2013a,Nanda2017}, however critical current alone does not offer direct information of the ABS spectrum. On the other hand, ABS spectroscopy requires either weakly coupled tunneling probes or microwave spectroscopy techniques, both challenging to combine with standard transport measurements. As a result, direct visualization of ABSs could only be achieved for a limited selection of material systems, such as atomic break junctions~\cite{Bretheau2013} carbon nanotubes~\cite{Pillet2010}, graphene~\cite{Bretheau2017}, and semiconductor nanowires~\cite{Chang2013,Tosi2019}, and not in combination with supercurrent measurements.

In this Letter we demonstrate concomitant measurement of phase-dependent critical current and Andreev bound state spectrum in a highly transmissive InAs JJ embedded in a DC SQUID loop. Tunneling spectroscopy reveals ABSs with near unity transmission probability. A non-sinusoidal CPR is derived from the ABS spectrum, showing excellent agreement with the one extracted from the SQUID critical current. Both measurements are reconciled in a short junction model where multiple ABSs, with various transmission probabilities, contribute to the entire supercurrent flowing in the junction. This work highlights hybrid planar JJs as ideal test grounds for emerging paradigms in condensed matter physics, such as ABS manipulation in highly transmitting devices~\cite{Zazunov2003,Janvier2015} and phase-tuned topology in multi-terminal geometries~\cite{vanHeck2014, Riwar2016,Pankratova2018}.

Planar JJs were defined in an InAs quantum well contacted by a thin epitaxial Al film grown \textit{in-situ}~\cite{Shabani2015,Kjaergaard2016}. Two lateral InAs regions covered by epitaxial Al constitute the superconducting leads, while the normal region is defined by selectively removing a stripe of epitaxial Al, exposing the semiconductor underneath. An energy diagram of the JJ under study is depicted in Fig.~\ref{fig:1}a. The strong superconductor-semiconductor coupling in the leads results in an induced superconducting DOS with gap $\Delta^*$, only slightly lower than the Al gap $\Delta$~\cite{Kjaergaard2017}. For a JJ much shorter than the superconducting coherence length, a well-known relation exists between the energy $E_i$ of an ABS and the phase difference between superconducting leads $\varphi$~\cite{Beenakker1991}:
\begin{equation}
E_i(\varphi)=\pm\Delta^*\sqrt{1-\tau_i\left[\sin\left(\varphi /2\right)\right]^2}
\label{eq:1}
\end{equation}
where $\tau_i$ is the transmission probability. The spectrum is periodic in $\varphi$ and approaches zero energy for $\tau_i \rightarrow 1$ and $\varphi=\pi$. Each populated ABSs carries a supercurrent
\begin{equation}
I_i(\varphi)=-\frac{2e}{h} \frac{\partial E_i}{\partial \varphi} .
\label{eq:2}
\end{equation}
Summing the current contribution of all ABSs results in the CPR of the junction, so that a junction with highly transmissive modes is characterized by a forward-skewed sine-like CPR.

Phase tuning was achieved by embedding the junction of interest, referred to as JJ1, in a superconducting loop (also made of epitaxial Al on InAs) together with a second InAs/Al JJ, named JJ2. Dimensions were chosen so that the critical current of JJ2 ($I_{\mathrm{C2}}$) was much larger than that of JJ1 ($I_{\mathrm{C1}}$). A superconducting tunneling probe with tunable transmission was integrated laterally to JJ1, allowing spectroscopy. Figure~\ref{fig:1}b shows a schematics of the device, while a false-colored scanning electron micrograph is shown in Fig.~\ref{fig:1}c. Figure~\ref{fig:1}d shows a zoom-in of JJ1 and the tunneling probe.

The device, shown in Fig.~\ref{fig:1}c, was defined by wet etching of the superconductor/semiconductor stack down to the insulating substrate. The epitaxial Al film was locally removed in order to form the normal regions of JJ1, JJ2 and the tunneling probe. The sample was subsequently covered by a $18~\mathrm{nm}$ thin film of insulating $\mathrm{HfO_2}$ and patterned with top gates. Junctions had dimensions $W_1=80~\mathrm{nm}$, $L_1=1.6~\mathrm{\mu m}$, $W_2=40~\mathrm{nm}$ and $L_2=5~\mathrm{\mu m}$, where $W$ is the separation between leads and $L$ their length. The loop was defined by a $160~\mathrm{nm}$ wide epitaxial Al stripe and encloses an area of $7~\mathrm{\mu m^2}$.

If not explicitly stated, the device was operated with gate voltages $V_{\mathrm{G1}}=V_{\mathrm{G2}}=0$. Additional gates, set to negative voltages, laterally confine electrons underneath the Al leads. For JJ2, this is achieved with gate $G_3$ while for JJ1 with two gates that also define a constriction in the 2DEG and are operated at the same gate voltage $V_{\mathrm{QPC}}$. This concept is presented in Fig.~\ref{fig:1}e, showing a cross section of JJ1 with the resulting conducting InAs region indicated by the shaded red area.

Devices were measured in a dilution refrigerator at a base temperature of $30~\mathrm{mK}$ by low-frequency lock-in techniques. The measurement setup is schematically shown in Fig.~\ref{fig:1}b. The tunneling differential conductance $dI_1/dV_1$ was measured by applying a voltage bias $V_{\mathrm{T}}=V_{\mathrm{DC}}+V_{\mathrm{AC}}$ to the tunneling probe via a low impedance IV converter and recording the resulting current $I_1$ and differential voltage $V_1$ (where $V_{\mathrm{DC}}$ was a DC signal and $V_{\mathrm{AC}}$, $V_1$ and $I_1$ had frequency $f_1$). The loop differential resistance $dV_2/dI_2$ was obtained by injecting a current $I_{\mathrm{IN}}=I_0+I_2$ in the loop and recording the differential voltage $V_2$ (where $I_0$ was a DC signal and $I_2$ and $V_2$ had frequency $f_2$). Measurements shown here were obtained with $I_2=2~\mathrm{nA}$, $V_{\mathrm{AC}}=3~\mathrm{\mu V}$, $f_1=131~\mathrm{Hz}$ and $f_2=113~\mathrm{Hz}$.

\begin{figure}
\includegraphics[width=\columnwidth]{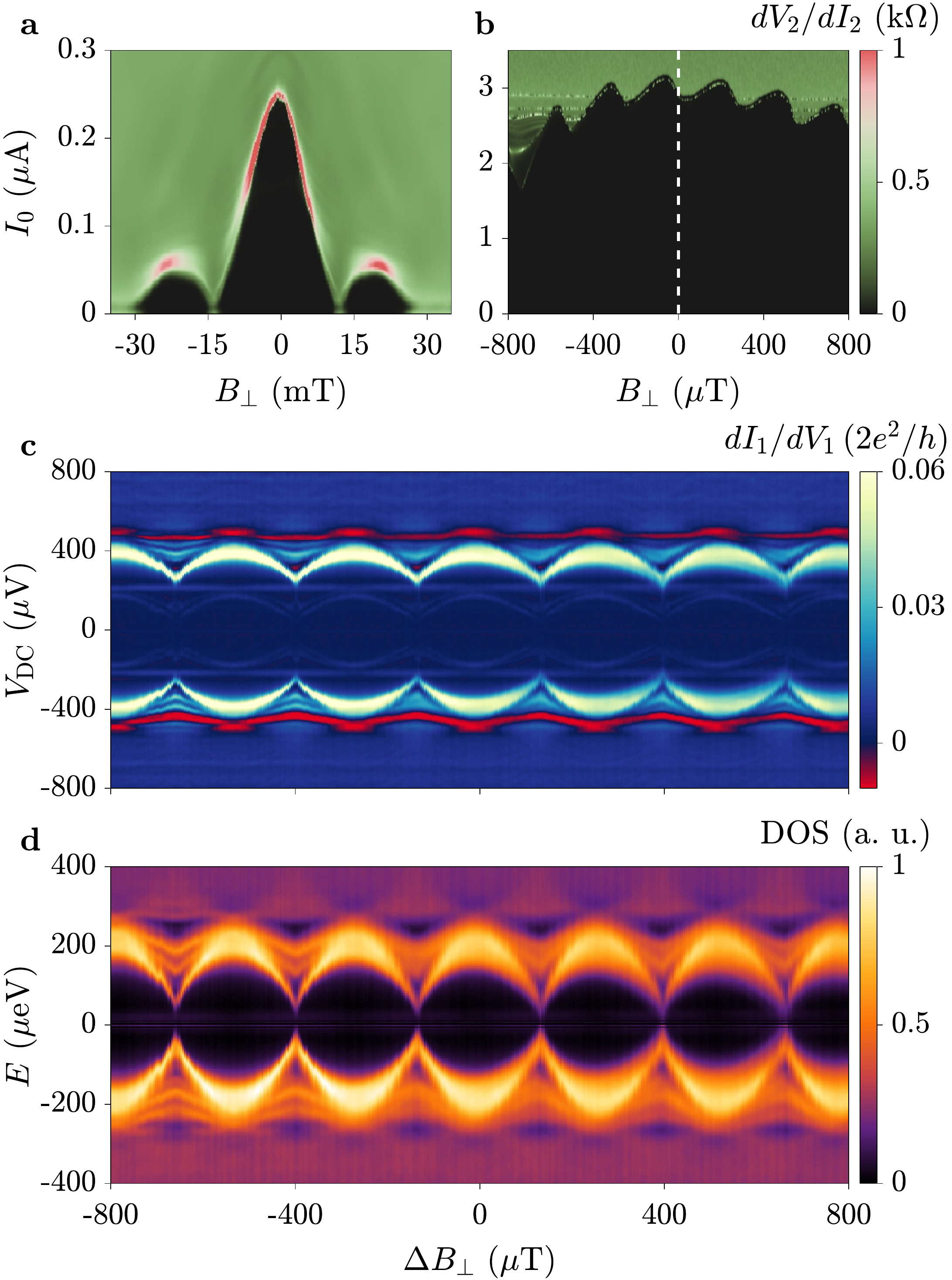}
\caption{(a) Loop resistance $dV_2/dI_2$ as a function of out-of-plane magnetic field $B_{\perp}$ and DC current $I_0$. Gate voltage $V_{\mathrm{G2}}$ was set very negative, so that a current can only flow in JJ1. (b) As in (a), but with the supercurrent flowing in both junctions. Notice the different magnetic field scale, highlighting supercurrent oscillations with a characteristic period consistent with one flux quantum impinging the loop area. The vertical dashed line marks the point with zero external magnetic field $B_{\perp}$. (c) Tunneling conductance $dI_1/dV_1$ as a function of bias voltage $V_{\mathrm{DC}}$, measured for the same magnetic field range as in (b). (d) Density of states within the normal region of JJ1, computed from the data in (c) using a deconvolution algorithm}.
\label{fig:2}
\end{figure}

Figure~\ref{fig:2}a shows the loop resistance $dV_2/dI_2$ as a function of out-of-plane magnetic field $B_{\perp}$ and $I_0$ for $V_{\mathrm{G2}}=-3~\mathrm{V}$, a situation in which JJ2 was closed and the current entirely flowed through JJ1. The critical current showed a Fraunhofer-like pattern, with maximum $I_{\mathrm{C1}}=240~\mathrm{nA}$ and zeros for $B_\perp=\pm 14~\mathrm{mT}$, consistent with one flux quantum $\Phi_0=h/(2e)$ impinging through the normal region of JJ1. Setting $V_{\mathrm{G2}}=0$ allowed the current to flow in both junctions. In this gate configuration, the critical current had mean value $I_{\mathrm{C2}}=2.9~\mathrm{\mu A}$ and was periodically modulated as a function of $B_{\perp}$ (see Fig.~\ref{fig:2}b). The modulation had amplitude $I_{\mathrm{C1}}$ and periodicity $265~\mathrm{\mu T}$, consistent with magnetic flux quanta impinging through the SQUID loop. The dashed vertical line in Fig.~\ref{fig:2}b marks the position of $B_{\perp}=0$, carefully measured as described in the Supplementary Information section. In this unbalanced SQUID regime ($I_{\mathrm{C2}}/I_{\mathrm{C1}}=12$), the critical current modulations are reminiscent of the CPR of JJ1~\cite{Nanda2017}, as discussed below.

The tunneling differential conductance $dI_1/dV_1$, measured in the same regime as in Fig.~\ref{fig:2}b, is shown in Fig.~\ref{fig:2}c. Consistent with a superconductor-insulator-superconductor junction, data showed a transport gap within $V_{\mathrm{SD}}=\pm2\Delta^*/e\approx400~\mathrm{\mu V}$ and regions of negative differential conductance. The energy spectrum of JJ1, shown in Fig.~\ref{fig:2}d, was obtained by applying a deconvolution algorithm \cite{Pillet2010} to the data of Fig.~\ref{fig:2}c. For this procedure, we assumed the tunneling probe was characterized by a superconducting DOS with gap $\Delta^*=200~\mathrm{\mu eV}$ and Dynes parameter $\gamma=0.02$ \cite{Kjaergaard2016,Suominen2017}. The deconvolved DOS displays a gap $\Delta^*=200~\mathrm{\mu eV}$ in which discrete and periodically modulated ABSs coexist. Some approach zero energy, indicating very high transmission, while others, with lower transmission, evolve closer to the gap edge.

\begin{figure*}
\includegraphics[width=\textwidth]{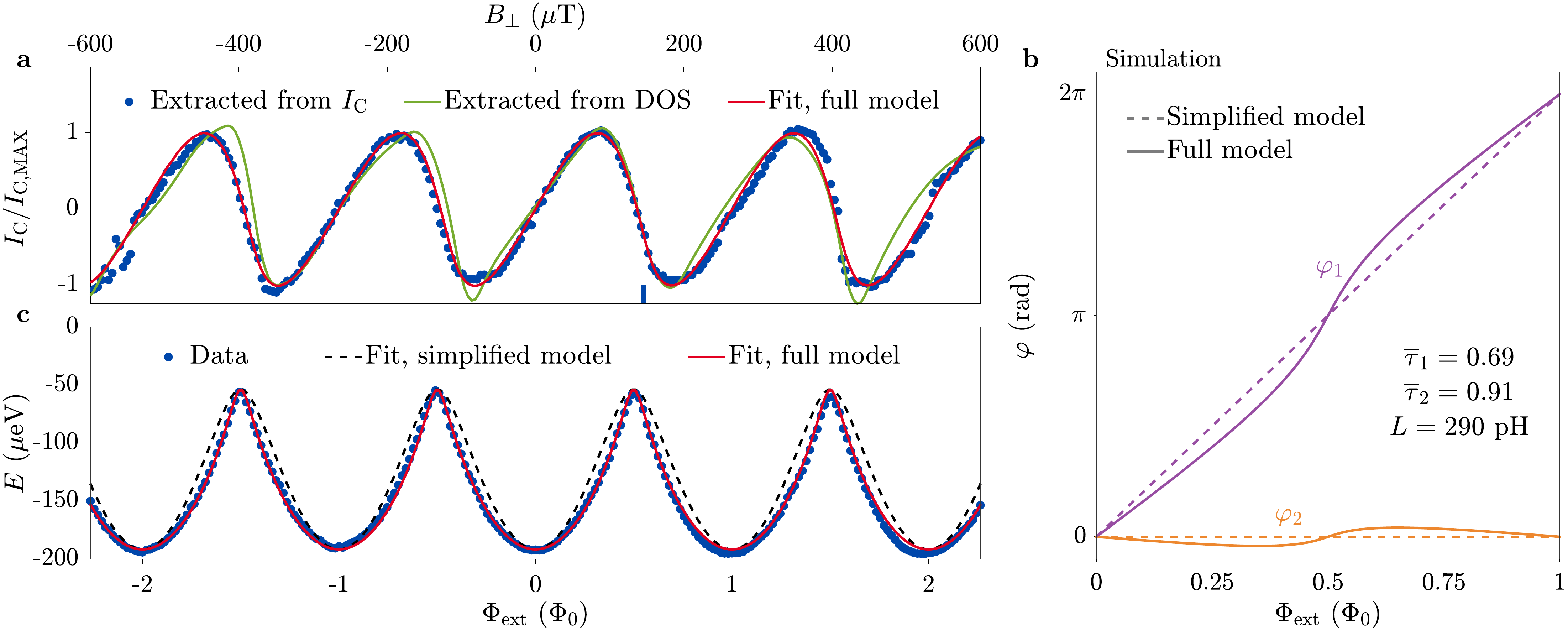}
\caption{(a) SQUID critical current as a function of out-of-plane magnetic field $B_{\perp}$ (blue dots, extracted from Fig.~\ref{fig:2}b) together with the critical current calculated based on the data in Fig.~\ref{fig:2}(d) (green), and a fit to the full numerica model (red). The fit used the average transmissions $\overline{\tau}_1$ and $\overline{\tau}_2$ as free parameters. Blue data points have been horizontally shifted by $170~\mathrm{\mu T}$ in order to match the other curves. The vertical blue marker shows the position where $B_{\perp}=0$ occured in the raw data. (b) Phase differences $\varphi_1$ and $\varphi_2$ across JJ1 and JJ2, respectively. Dashed lines refer to the simplified model, which considers $I_{\mathrm{C2}}\gg I_{\mathrm{C1}}$ and no loop inductance. Solid lines are calculated using our full model and the fit result of (a). (c) Energy of the most transmissive ABSs visible in Fig.~\ref{fig:2}a (blue dots) together with a fit to Eq.~\ref{eq:1} that includes the non-linear mapping between $\Phi_{\mathrm{ext}}$ and $\varphi_1$ shown in (b) (red line). The dashed black line is a fit to Eq.~\ref{eq:1} using the simplified assumption $\varphi_1=2\pi\Phi_{\mathrm{ext}}$. In this case the fit was forced to match the data in the points of highest energy.}
\label{fig:3}
\end{figure*}

The derived DOS was then used to compute the field dependent supercurrent as the sum of the contributions of each occupied ABSs ($E_i<0$), as per Eq.~\ref{eq:2}. Here we used the approximation:
\begin{equation}
I_1(\varphi_1)=
-\frac{2e}{h}\sum_i \frac{\partial E_i}{\partial \varphi_1}
\approx-\frac{2e}{h}\int_{V_{\mathrm{DC}}}
\frac{\partial \mathrm{DOS}}{\partial\varphi_1}\epsilon d\epsilon
\label{eq:3}
\end{equation}
that is, we substituted the summation with an integral of the phase derivative of the DOS times energy, which is readily computed on the data of Fig.~\ref{fig:2}c. Figure~\ref{fig:3}a compares the field dependent supercurrent of JJ1 extracted from the critical current modulation of Fig.~\ref{fig:2}b (blue dots) and that numerically derived from the ABSs of Fig.~\ref{fig:2}c and Eq.~\ref{eq:3} (green line). The matching of the two curves is striking, confirming that indeed the CPR of JJ1 is completely determined by its ABS spectrum. Furthermore, the pronounced forward-skewness indicates a high average transmission. Points extracted from the critical current (blue dots) were shifted by $170~\mathrm{\mu T}$ with respect to the magnetic field they were measured at (the vertical blue marker indicates the position where $B_{\perp}=0$ was identified in Fig.~\ref{fig:2}b). As discussed in the following, the shift is due to current-induced magnetic fluxes induced within the device and is accounted for by our model.

In a first simplified approximation, an external flux $\Phi_{\mathrm{ext}}$ through the loop results in a phase difference $\varphi_1\approx 2\pi\Phi_{\mathrm{ext}}$ across JJ1 while the phase difference $\varphi_2$ across JJ2 remains zero, as shown by dashed lines in Fig.~\ref{fig:3}b. However, in the present case two complications arise. First, JJ1 is in parallel with a second highly transmissive JJ, which introduces a non-linear Josephson inductance. Second, the epitaxial Al used for the superconducting loop is particularly thin, resulting in a finite kinetic inductance $L_{\mathrm{K}}\approx290~\mathrm{pH}$ (see Supplementary Information). In this framework, the effective flux $\Phi$ impinging through the loop is the difference between the externally applied flux $\Phi_\mathrm{ext}$ and the current induced flux $L_{\mathrm{K}}(I_2-I_1)/2$. The JJs are both in the ballistic regime, resulting in non-sinusoidal CPRs, which depends on their average transmissions $\overline{\tau}_1$ and $\overline{\tau}_2$ (see Eq.~\ref{eq:2}):
\begin{equation}
\Phi=\Phi_{\mathrm{ext}}-L_{\mathrm{K}}\left[ I_2(\varphi_2,\overline{\tau}_2)-I_1(\varphi_1,\overline{\tau}_1)\right]/2.
\label{eq:M_2}
\end{equation}
Taking into account flux quantization, we substitute $\varphi_2=\varphi_1-2\pi\Phi$, resulting in an equation of two unknowns, $\varphi_1$ and $\Phi$, that is solved self-consistently to obtain $\Phi_{\mathrm{ext}}(\varphi_1,\Phi)$ and $I_\mathrm{SQUID}(\varphi_1,\Phi)=I_1(\varphi_1,\tau_1)+I_2(\varphi_1-2\pi\Phi,\tau_2)$. Solutions are further constrained using physical arguments: for each value of $\Phi_{\mathrm{ext}}$, the SQUID critical current is the maximum of $I_\mathrm{SQUID}$. Similarly, the situation with only circulating currents (equivalent to the case where tunneling spectroscopy is measured) is obtained for $I_\mathrm{SQUID}=0$.

Fitting of our full model to the critical current of Fig.~\ref{fig:2}b resulted in average transmissions $\overline{\tau}_1=0.69$ and $\overline{\tau}_2=0.92$, consistent with JJ1 having a larger electrodes separation than JJ2. The fit is shown in Fig.~\ref{fig:3}a (red line), with the computed $\varphi_1$ and $\varphi_2$ as a function of $\Phi_{\mathrm{ext}}$ plotted in Fig.~\ref{fig:3}b (solid lines). Deviations from the simplified model are particularly relevant close to $\Phi_{\mathrm{ext}}=\Phi_0/2$, where non-linearities occur. The horizontal shift observed between the measured critical current of Fig.~\ref{fig:2}b and that numerically computed of Fig.~\ref{fig:3}a (indicated by the blue mark) is understood as due to the induced flux in the loop, driven by the large circulating current present in the measurement of Fig.~\ref{fig:2}b and absent when performing tunneling spectroscopy (Fig.~\ref{fig:2}c). The sensitivity of the calculated SQUID critical current on the choice of input parameters is presented in the Supplementary Information.

We now provide a quantitative understanding of the measured ABSs. Figure~\ref{fig:3}c shows the energy of the more transmissive ABSs extracted from Fig.~\ref{fig:2}d (blue dots) together with a fit of Eq.~\ref{eq:1} that takes into account the non-linear relation between $\Phi_\mathrm{ext}$ and $\varphi_1$ represented in Fig.~\ref{fig:3}b (purple line). The fit of our full model finds excellent agreement to the measured ABS and results in $\tau_1=0.92$, among the highest values ever reported. The states with lower transmission visible in Fig.~\ref{fig:2}b were fit with transmission parameters $0.54$ and $0.15$. We contrast this with a fit using the simplified mapping between $\varphi$ and $\Phi_{\mathrm{ext}}$ (black dashed line), which does not take into account the presence of JJ2 and the loop kinetic inductance. The low quality of this fit highlights how device complexities might result in severe distortions of the measured ABSs and should be properly accounted for. 

In conclusion, we measured highly transmissive planar JJs embedded in an InAs/Al hybrid heterostructure. Thanks to a novel device design, we related the junction ABS spectrum to its CPR, finding remarkable agreement. Supercurrent gave an average transmission $\overline{\tau}_1=0.69$, resulting in a forward-skewed CPR. On the other hand, tunneling spectroscopy provided a much richer description of the device, revealing sets of discrete ABSs populating the superconducting gap. We measured ABSs with almost unity transmission, well separated from other states and the continuum, and energy approaching zero for $\varphi_1=\pi$. These levels are a promising starting point for engineering topological states either via phase biasing in multi-terminal geometries~\cite{vanHeck2014,Riwar2016} or in combination with small Zeeman fields~\cite{Pientka2017,Hell2017}. Our work also highlights the complexities inherent in SQUID devices with ballistic junctions and large kinetic inductance, and provides a framework for future studies.

\begin{acknowledgments}
This work was supported by Microsoft Corporation, the Danish National Research Foundation, and the Villum Foundation. F.~Nichele acknowledges support from European Research Commission, grant number 804273. We thank A.~Stern, E.~Berg, F.~Setiawan, A.~Keselman, H.~Pothier, L.~Bretheau and P.~Scarlino for fruitful discussions.
\end{acknowledgments}

\bibliography{Bibliography}

\section*{Supplementary Information}
\setcounter{figure}{0}
\setcounter{equation}{0}
\setcounter{table}{0}
\renewcommand{\thefigure}{S.\arabic{figure}}
\renewcommand{\theequation}{S.\arabic{equation}}

\begin{figure}
\includegraphics[width=\columnwidth]{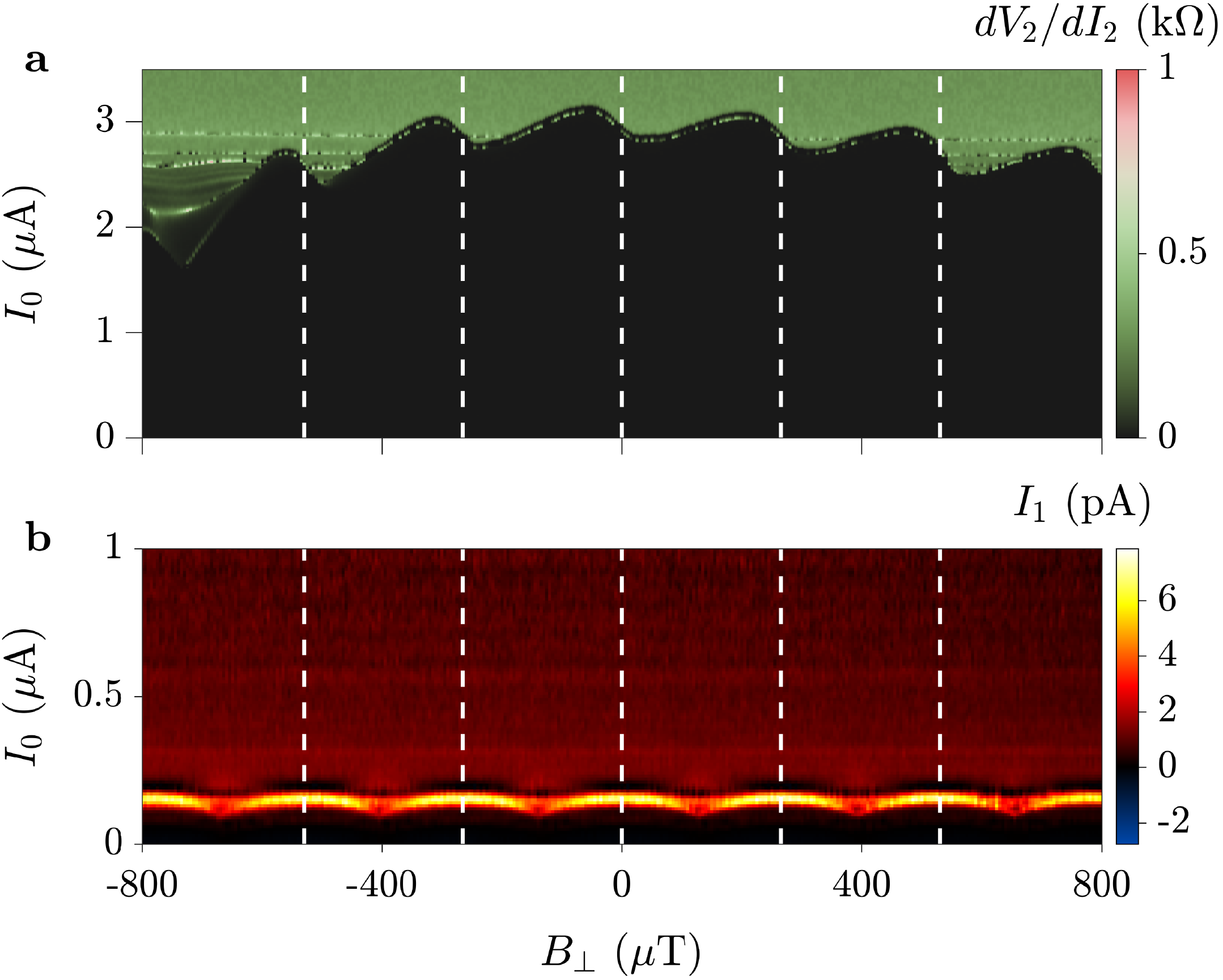}
\caption{\textbf{Field offset compensation.} \textbf{a}, Loop resistance $dV_2/dI_2$ as a function of the corrected out-of-plane magnetic field $B_{\perp}$ and DC current $I_0$ as in Fig.~2b of the Main Text. \textbf{b}, Tunneling current $I_1$ recorded simultaneously to the measurement in \textbf{a}. Zero magnetic field has been fixed correspondingly to the maximum of the gap modulation. Vertical dashed lines in both \textbf{a} and \textbf{b} indicate integer multiples of the flux quantum $\Phi_0$.}
\label{fig:S_1}
\end{figure}

\begin{figure*}
\includegraphics[width=\textwidth]{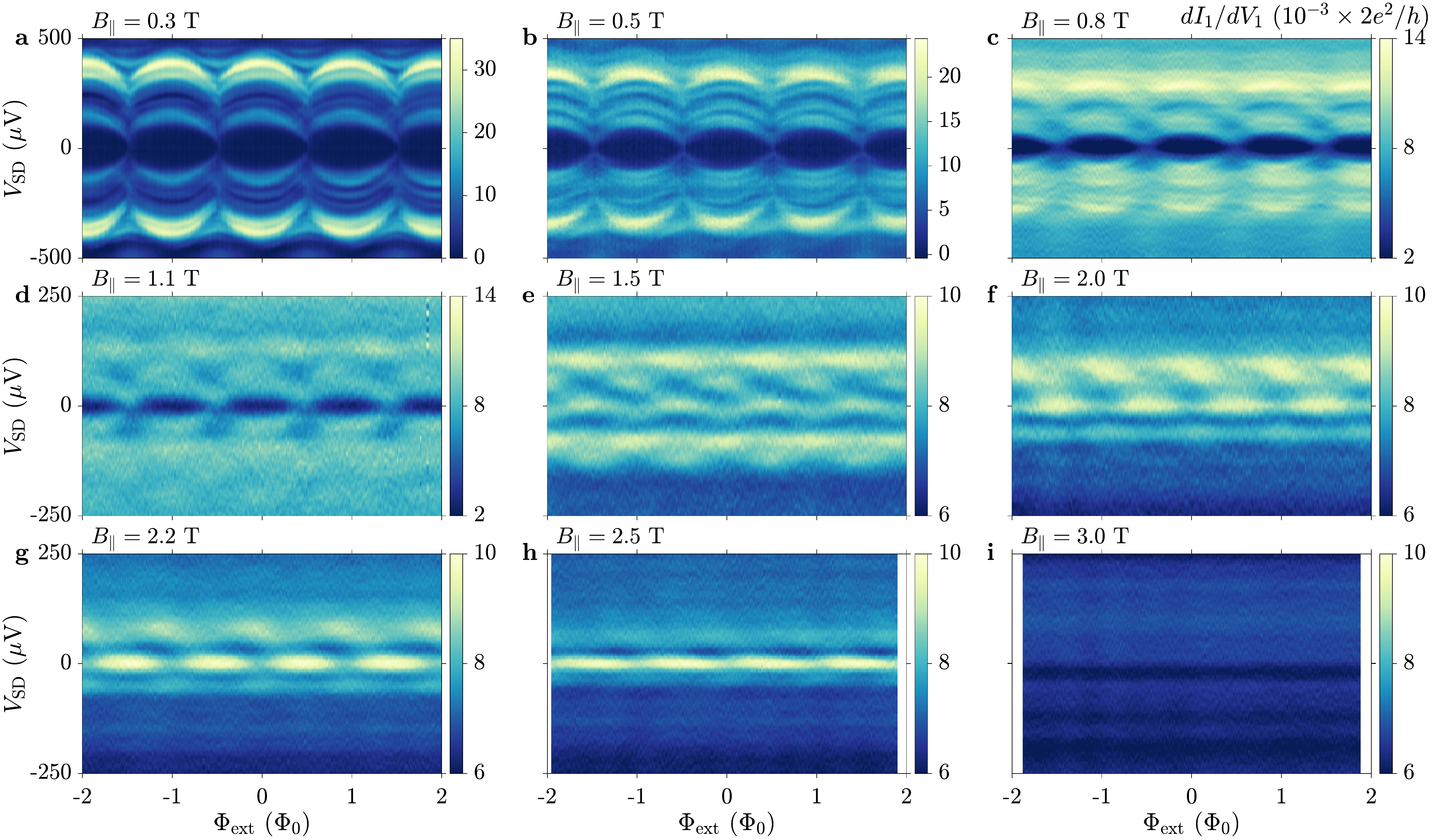}
\caption{\textbf{Tunneling conductance as a function of in-plane magnetic field.} Conductance $dI_1/dV_1$ of the tunneling probe measured as a function of an in-plane magnetic field oriented along the $y$ direction as defined in Fig.~1c of the Main Text. Despite a superconducting tunneling probe was used, a moderate magnetic field allows one to perform direct spectroscopy as with a normal probe (see text). In a large in-plane magnetic field, periodic ZBPs emerge for semi-integer values of the magnetic flux, interpreted as the overlap of multiple ABSs in proximity to a band inversion.}
\label{fig:S_2}
\end{figure*}

\subsection{Wafer structure}
Samples were fabricated from an heterostructure grown on an InP substrate by molecular beam epitaxy techniques. The active region enclosed, from bottom to top, of a $25~\mathrm{nm}$ $\mathrm{In_{81}Al_{19}As}$ layer, a $4~\mathrm{nm}$ $\mathrm{In_{81}Ga_{19}As}$ bottom barrier, a $5~\mathrm{nm}$ $\mathrm{InAs}$ quantum well, a $5~\mathrm{nm}$ $\mathrm{In_{90}Al_{10}As}$ top barrier, two $\mathrm{GaAs}$ monolayers and a $8.7~\mathrm{nm}$ Al film. The Al layer was grown in the same chamber as the III-V heterostructure, without breaking vacuum in between~\cite{Shabani2015}. The InAs 2DEG has been characterized in a Hall bar geometry, where the Al film has been selectively removed before deposition of a $\mathrm{HfO_{2}}$ dielectric. The mobility peaked at $18.000~\mathrm{cm^2V^{-1}s^{-1}}$ for an electron density of $1.5\times10^{12}~\mathrm{cm^{-2}}$, corresponding to an electron mean free path of $360~\mathrm{nm}$. Sample fabrication has been extensively discussed in Ref.~\onlinecite{Fornieri2019}.

The kinetic inductance of the superconducting loop was estimated from the loop geometry and the normal state resistivity of the heterostructure stack as~\cite{Annunziata2010}:
\begin{equation}
L_\mathrm{K}=\frac{l}{w}\frac{h}{2\pi^2}\frac{R_{\square}}{\Delta}=290~\mathrm{pH},
\label{eq:S_0}
\end{equation}
where $l$ and $w$ are the length and width of the superconducting stripe defining the SQUID loop and $R_{\square}$ is the normal state resistivity of the Al/InAs heterostructure.
In the present case $l/w=48$, while $R_{\square}$ was separately measured above the Al critical temperature in a Hall bar geometry where the top Al film was not removed, giving $6.4~\mathrm{\Omega}$. In comparison, the geometric inductance of the loop was estimated in $4~\mathrm{pH}$ only.
In the Supplementary Information section, fitting of the data of Fig.~3 of the Main Text was repeated using different values of $L_{\mathrm{K}}$. Meaningful results in terms of transmission coefficients could be obtained only for $250~\mathrm{pH}\leq L_{\mathrm{K}}\leq350~\mathrm{pH}$.

\subsection{Experimental identification of zero out-of-plane magnetic field}
Superconducting vector magnets are typically characterized by hysteresis and magnetic field offsets of the order of a few $\mathrm{mT}$, much larger than the periodicity of the supercurrent modulation measured in this work ($265~\mathrm{\mu T}$). As the fitting procedure presented in the main text crucially relies on the correct identification of the $B_{\perp}=0$ condition, particular care was taken to properly account for this experimental uncertainty.

For this reason, while measuring the loop resistance $dV_2/I_2$, the AC current $I_1$ flowing through the high impedance tunneling probe was recorded via a low impedance IV converter. In Fig.~\ref{fig:S_1} we compare the loop resistance $dV_2/I_2$ (as in Fig.~2b of the Main Text), with the \textit{simultaneously} measured current $I_1$ flowing through the tunneling probe. The raising DC current $I_0$ injected through the loop resulted in an increasing DC voltage bias between the two sides of the tunneling probe, similarly to the measurement of Fig.~2c of the Main Text. With the loop in the superconducting regime, the DC voltage bias across the tunneling junction is given by $I_0$ times the resistance to ground from the bottom contact of the SQUID loop. Periodic modulation of the superconducting gap are clearly visible in Fig.~\ref{fig:S_1}b. We set $B_{\perp}=0$ to the point of maximum gap size in Fig.~\ref{fig:S_1}b, concomitant to the maximum of the slowly varying envelop of the SQUID critical current visible in Fig.~\ref{fig:S_1}a. Dashed vertical lines mark the magnetic field periodicity, visible in both critical current and tunneling spectrum.
The CPR of a JJ is known to be sine-like (with an eventual forward skewness) and have zero current for zero phase difference. In the measurement of Fig.~\ref{fig:S_1}a the point identified as $B_{\perp}=0$ does not occur at a zero of the forward-skewed sinusoidal modulation of critical current. This anomalous flux offset, which should not be confused for a phase offset, is explained in our model as due to the large induced flux in the loop when $I_0\approx I_{\mathrm{C2}}$.

\begin{figure}
\includegraphics[width=\columnwidth]{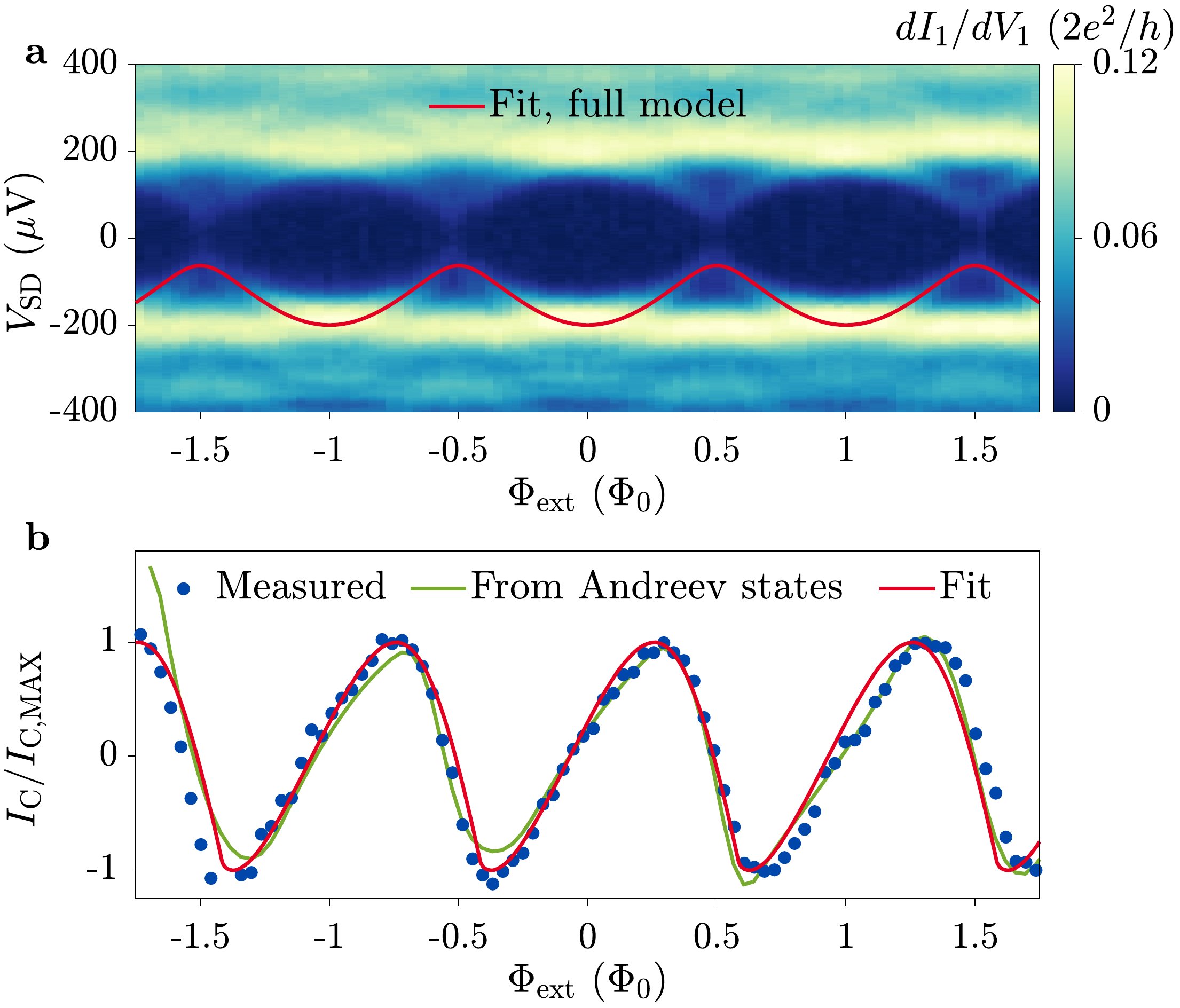}
\caption{\textbf{Device with normal tunneling probe.} \textbf{a}, Tunneling conductance $dI_1/dV_1$ as for Fig.~2b of the Main Text, but for a device with a normal tunneling probe. In this case, the conductance is directly proportional to the ABS spectrum, without requiring numerical deconvolution. The red line is a fit of the ABS, yielding a transmission $\tau_1=0.90$. \textbf{b}, Measured critical current modulation of the SQUID loop (blue dots) together with that derived from the ABS spectrum in \textbf{a} (green line) and a fit to the numerical full model, resulting an average transmission $\tau_1=0.7$.}
\label{fig:S_3}
\end{figure}

\subsection{In-plane magnetic field dependence}
The sample presented in the Main Text has been further characterized in terms of tunneling spectroscopy as a function of an in-plane magnetic field $B_{\parallel}$, applied perpendicular to the current direction in the JJs ($y$ direction in Fig.~1c of the Main Text). Results are summarized in Fig.~\ref{fig:S_2}. A moderate in-plane magnetic field ($B_{\parallel}\geq200~\mathrm{mT}$) induces a finite zero energy density of states (DOS) underneath the large Al plane beyond the tunneling junction, while leaving the smaller Al segments constituting the leads of the JJs still in a hard-gap regime. As a result, transport occurs also at zero bias and the measured $dI_1/dV_1$ signal is directly proportional to the DOS within the normal region of the junction, without the need of numerical deconvolution. This phenomenon has been characterized, for a similar setup, in Ref.~\onlinecite{Suominen2017}. Figure~\ref{fig:S_2}a shows the situation for $B_{\parallel}=300~\mathrm{mT}$, where the effect of Zeeman field on the ABSs is still modest. The ABSs energy closely approaches zero energy, providing further evidence of their very high transmission and consistent with the deconvolved DOS of Fig.~2c of the Main Text.

As the magnetic field is further increased, a gradual closure of the superconducting gap was observed, together with the onset of zero bias conductance peaks (ZBPs) beyond $1.5~\mathrm{T}$ for semi-integer values of the flux quantum (see Fig.~\ref{fig:S_2}f). Increasing the field further, ZBPs spread over the full phase space (see Fig.~\ref{fig:S_2}h). Such a behavior resembles that expected for a topological transition in a phase controlled JJ, with Majorana zero modes existing at the junction edges \cite{Hell2017,Pientka2017}. Evidence supporting the existence of such transition were recently reported by some of us in Ref.~\onlinecite{Fornieri2019}, using a device very similar to the one presented here. However, two noticeable differences appear between the ZBPs observed here and those observed in Ref.~\onlinecite{Fornieri2019}. First, in Ref.~\onlinecite{Fornieri2019} spectroscopy of ZBPs revealed a discrete nature. As a function of magnetic field, the ZBPs gradually evolved from the high energy continuum to zero energy, before starting oscillating at even higher fields. These observations are consistent with a Majorana zero mode interpretation. In contrast, ZBPs observed here are broad in energy, present no fine structure and do not oscillate in energy as a function of magnetic field. Second, in Ref.~\onlinecite{Fornieri2019}, locating ZBPs consistent with Majorana modes required careful tuning of the chemical potential, as expected for Majorana modes in finite-size devices \cite{Setiawan2019}. In the present case, ZBPs are found without the need of gate voltage tuning, suggesting a more frequent occurrence. We speculate the observed ZBPs are related to those recently reported in Ref.~\onlinecite{Ren2019} and interpreted as bands of highly transmissive ABSs crossing zero energy in the proximity of a field driven band inversion. Further work is needed to better elucidate the nature of the ZBPs in Fig.~\ref{fig:S_2}.

\subsection{Sample with normal tunneling probe}
Andreev bound states with transmission approaching unity have been consistently measured, in this work, over several devices. The use of a superconducting tunneling probe generally results in sharp spectroscopic data, however direct visualization of the spectrum requires numerical deconvolution~\cite{Pillet2010}. Here we present measurements of a device similar to that of Fig.~1c of the Main Text, but with the Al film beyond the tunneling probe removed, allowing direct visualization of the DOS. With respect to the device of the Main Text, the one presented here was fabricated on the same heterostructure as Ref.~\onlinecite{Fornieri2019}, JJ1 had a width $W_1=40~\mathrm{nm}$ and the critical currents of JJ1 and JJ2 were $I_{\mathrm{C1}}=140~\mathrm{nA}$ and $I_{\mathrm{C2}}=1.3~\mathrm{\mu A}$ respectively.
Tunneling conductance $dI_1/dV_1$ is shown in Fig.~\ref{fig:S_3}a and qualitatively agrees with the deconvolved DOS of Fig.~2c of the Main Text. The highly transmissive ABSs were fit with our numerical model, yielding $\tau_1=0.90$. In Fig.~\ref{fig:S_3}b we compare the measured SQUID critical current modulation (blue dots) with the CPR of JJ1 computed using Eq.~3 in the Main Text (green line), finding excellent agreement. Fitting of the measured critical current modulation (red line) results in an average transmission of JJ1 of $\overline{\tau}_1=0.7$.

\subsection{Sensitivity of the numerical model result on input parameters}
\begin{figure*}
\includegraphics[width=\textwidth]{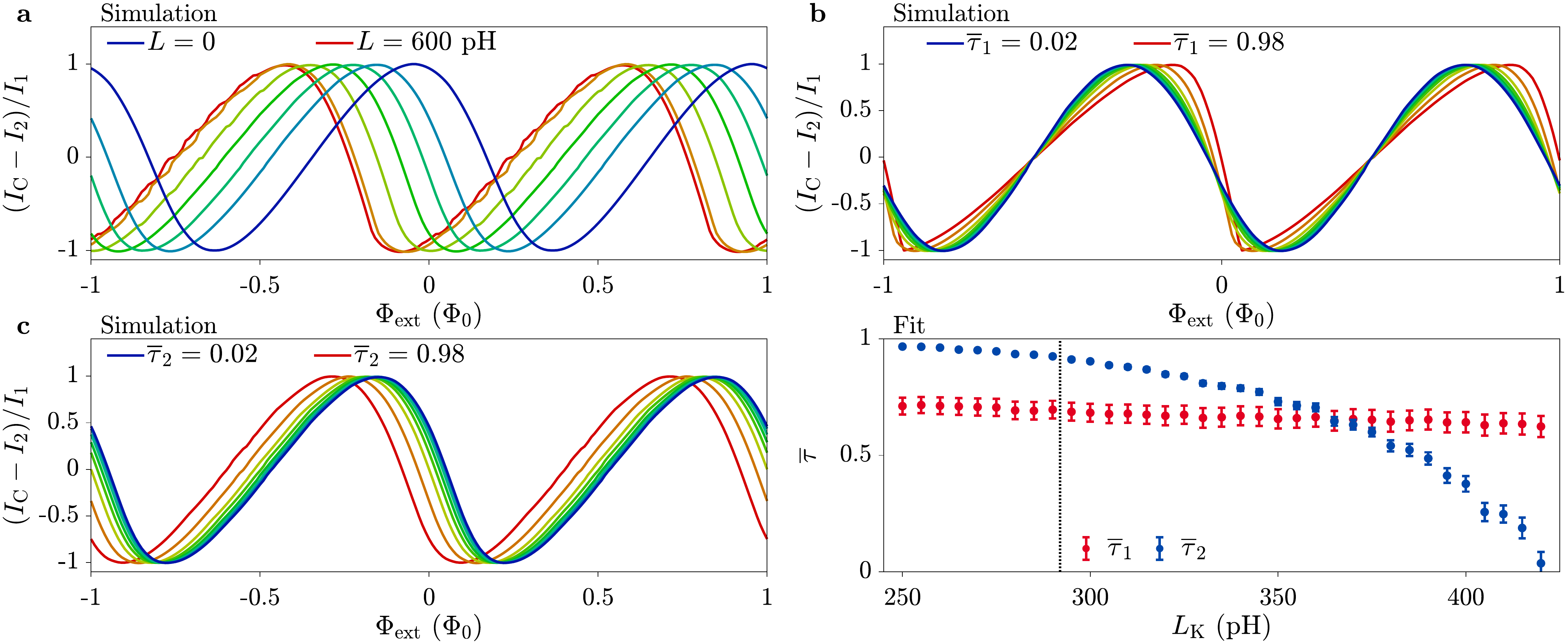}
\caption{\textbf{Sensitivity of the numerical model on input parameters.} SQUID critical current modulation calculated according to Eq.~6 as a function of: \textbf{a}, Kinetic inductance, with $L_{\mathrm{K}}=0,~175,~275,~375,~475,~575,~600~\mathrm{pH}$ \textbf{b}, Average transmission of JJ1, with $\overline{\tau}_1=0.98,~0.90,~0.74,~0.58,~0.42,~0.26,~0.10,~0.02$. \textbf{c}, Averange transmission of JJ2, with $\overline{\tau}_2=0.98,~0.90,~0.74,~0.58,~0.42,~0.26,~0.10,~0.02$. If not explicitly stated, the same parameters as in Fig.~3b of the Main Text were used. The shape of the critical current modulation if mostly affected by changes in $\overline{\tau}_1$, while changes in $\overline{\tau}_2$ and $L_{\mathrm{K}}$ result in a shift of the curve as a function of externally applied flux. \textbf{d}, Fit results for the critical current modulation shown in Fig.~3a of the Main Text with $\overline{\tau}_1$ and $\overline{\tau}_2$ as fit parameters and as a function of the chosen value of loop kinetic inductance $L_{\mathrm{K}}$. The dashed vertical line indicates the value of kinetic inductance estimated for the heterostructure in use.}
\label{fig:S_4}
\end{figure*}

In Fig.~\ref{fig:S_4} we plot the results of the numerical model used in the Main Text varying, one at a time, the input values of the unknown parameters: $L_{\mathrm{K}}$ (Fig.~\ref{fig:S_4}a), $\overline{\tau}_1$ (Fig.~\ref{fig:S_4}b), and $\overline{\tau}_2$ (Fig.~\ref{fig:S_4}c). As expected for an unbalanced SQUID, the CPR skewness is mainly determined by $\overline{\tau}_1$, while $\overline{\tau}_2$ and $L_{\mathrm{K}}$ give rise to a horizontal shift of the SQUID critical current plotted against $\Phi_{\mathrm{ext}}$. In the Main Text, we performed a fit of the SQUID critical current as a function of externally applied flux (Fig.~3a of the Main Text) with $\overline{\tau}_1$ and $\overline{\tau}_2$ as fit parameters and kinetic inductance $L_{\mathrm{K}}=290~\mathrm{pH}$ derived from the heterostructure resistivity and device geometry (see Methods). In Fig.~\ref{fig:S_4}d we show the result of the same fit for different input values of $L_{\mathrm{K}}$. We notice that meaningful results were obtained only for the interval of $L_{\mathrm{K}}$ values shown on the horizontal axis. As expected for an unbalanced SQUID, $\overline{\tau}_1$ is largely unaffected by the choice of $L_{\mathrm{K}}$. In contrast, $\overline{\tau}_2$ strongly depends on the input choice for $L_{\mathrm{K}}$, as both parameters mainly give rise to a flux offset of the critical current. The separation between superconducting electrodes in the JJs under study is $80~\mathrm{nm}$ for JJ1 and $40~\mathrm{nm}$ for JJ2, respectively. It is therefore reasonable to assume $\overline{\tau}_2>\overline{\tau}_1$, restricting the acceptable values of $L_{\mathrm{K}}$ between $250$ and $350~\mathrm{pH}$. This interval is in excellent agreement with the independently calculated value $L_{\mathrm{K}}=290~\mathrm{pH}$ used in the Main Text (see dashed vertical line in Fig.~\ref{fig:S_4}d). Other parameters entering the calculation are the critical currents of the junctions $I_{\mathrm{C1,2}}$ and the induced superconducting gap $\Delta^*$, all derived from experimental data.

\end{document}